\documentclass[aps,pre,twocolumn,letterpaper,showpacs,floatfix,superscriptaddress]{revtex4}
\usepackage{graphicx}

\begin{document}

\title{The random growth of interfaces as a subordinated process}

\author{R.~Failla}
\affiliation{Center for Nonlinear Science, University of North Texas, P.O. Box 311427, Denton, Texas 76203-1427, USA}
\author{P.~Grigolini}
\email{grigo@unt.edu}
\affiliation{Center for Nonlinear Science, University of North Texas, P.O. Box 311427, Denton, Texas 76203-1427, USA}
\affiliation{Istituto dei Processi Chimico Fisici del CNR, Area della Ricerca di Pisa, Via G. Moruzzi, 56124, Pisa, Italy}
\affiliation{Dipartimento di Fisica dell'Universita' di Pisa and INFM, via Buonarroti 2, 56127 Pisa, Italy}
\author{M.~Ignaccolo}
\affiliation{Center for Nonlinear Science, University of North Texas, P.O. Box 311427, Denton, Texas 76203-1427, USA}
\author{A.~Schwettmann}
\affiliation{Center for Nonlinear Science, University of North Texas, P.O. Box 311427, Denton, Texas 76203-1427, USA}

\date{\today}

\begin{abstract}

We study the random growth of surfaces from within the perspective of a single column, namely, the fluctuation of the column height around the mean value, $y(t) \equiv h(t)-\left< h(t) \right>$, which is depicted as being subordinated to a standard fluctuation-dissipation process with friction $\gamma$.  We  argue that the main properties of Kardar-Parisi-Zhang theory, in one dimension, are derived by identifying the distribution of return times to $y(0) = 0$, which is a truncated inverse power law, with the distribution of subordination times. The agreement of the theoretical prediction with the numerical treatment of the $ 1 + 1$  dimensional model of ballistic deposition is remarkably good, in spite of the finite size effects affecting this model.

\end{abstract}


\maketitle
The random growth of surfaces is a subject of increasing interest: The number of citations of the pioneer paper  \cite{KPZ}, where the Kardar-Parisi-Zhang (KPZ) equation was originally proposed, at the moment of writing this paper is of 829 in the journals of American Physical Society alone. The subject is discussed in excellent review papers \cite{zhang} and books \cite{stanley,meakin}. The interest for this field is not limited to the nanotechnology applications (see Refs.~\cite{moro} and \cite{more} for recent examples). A simple model such as the Ballistic Deposition (BD) \cite{stanley} is an example of self-organization:  As pointed out by 
Family \cite{family}, a growing surface spontaneously evolving into a steady state with universal fractal properties is similar to the mechanism of self-organized criticality \cite{bak}.  The columns of the material growing due to the deposition of particles can be thought of as the individuals of a society. The joint action of the randomness driving the particle deposition and the interaction among columns results in the emergence of anomalous scaling coefficients, which can be interpreted as the signature of cooperation.
However, only little attention has been devoted so far to studying the dynamics of the single individuals of this society, namely, the single growing columns of the sample under study. Usually the authors of this field of research study the correlation among distinct columns \cite{chow} without paying attention to the dynamics of an individual. Yet, a single column is expected to carry information about cooperation.

The single column perspective was recently adopted by Merikoski \emph{et al.} \cite{merikoski}  to study  combustion fronts in paper. The individual property under observation is
\begin{equation}
\label{individual}
y(t) = h(t) - \left< h(t) \right>,
\end{equation}
where $h(t)$ denotes the height of a single column at time $t$ and $\left< h(t) \right>$ the average over the heights of the columns of the whole sample. The authors of Ref.~\cite{merikoski} record the times at which the variable $y(t)$ changes sign and build up the corresponding time series $t_{i}$ so as to create  the new time series $\tau_{i} = t_{i+1} - t_{i}$, namely, the set of time distances between two consecutive recrossings of the origin $y = 0$. The distribution density $\psi_{D}(\tau)$ is shown \cite{merikoski} to be an inverse power law with index $\mu_{D}$,  fulfilling the relation
\begin{equation}
\label{individualandsociety}
\beta = 2 - \mu_{D}.
\end{equation}
The coefficient $\beta$ refers to the interface growth prior to saturation, a physical condition where the standard deviation of all $L$ columns, the interface width
\begin{equation}
w(L,t)\equiv\sqrt{\frac{1}{L}\,\sum_{i=1}^{L}\, \left[h_i(t)-\left< h(t) \right>\right]^2},
\end{equation}
grows as $w(L,t) \propto t^{\beta}$. Eq.~(\ref{individualandsociety})  establishes a connection between a single column property, $\mu_{D}$, and a collective property, $\beta$, thereby playing an important role for the perspective adopted in this paper. The theoretical foundation for this important relation is given in earlier papers  \cite{dingdang,krugandobbs,searson,questioningthemsleves} and has been more recently discussed by Majumdar \cite{majumdar}.

In this paper we prove that the KPZ  condition emerges from the identification of  $\psi_{D}(\tau)$ with the distribution function $\psi_{S}(\tau)$,  the essential ingredient of the subordination theory
 \cite{sokolov,metzlerandklafter,barkaiandsilbey}  stemming from the original work of Montroll and Weiss \cite{montrollandweiss}.  In the subdiffusion case, anomalous diffusion is derived from the ordinary diffusion process by assuming that the time distance between one jump and the next is determined by the inverse power law time distribution $\psi_{S}(\tau)$ with the index $\mu < 2$. According to this theory
\begin{equation}
\label{traditional}
\beta = \frac{\mu_{S} -1}{2} \equiv \frac{\alpha}{2}.
\end{equation}
From the identification of $\mu_{S}$, Eq.(4), with $\mu_{D}$, Eq.(2), we obtain the anomalous scaling parameter $\beta = 1/3$, which is the KPZ prediction.

However, to prove that the KPZ condition is a subordinated process, we have to show that Eq.~(\ref{individualandsociety}) can be derived from the assumption that the times $\tau_{i}$ are not correlated, an essential property of $\psi_{S}(\tau)$. For this purpose, we assume that $y$ is a diffusion process with scaling,
\begin{equation}
\label{diffusionscaling}
p(y,t) = \frac{1}{t^{\beta}} F\left(\frac{y}{t^{\beta}}\right).
\end{equation}
The number of particles located in a strip of size $dy$ around $y = 0$, $N(t)$, is proportional to $p(0,t)dy$. Thus, from Eq.~(\ref{diffusionscaling}) we get
\begin{equation}
\label{particleatorigin}
N(t)  = \frac{A}{t^{\beta}} ,
\end{equation}
where $A$ is a constant proportional to $F(0)$. On the other hand, in the scaling regime the particles that are found at the origin at a given time $t$ are only the particles that went back as a consequence of one or more recurrences. For this reason we write
\begin{equation}
\label{recurrences}
N(t) \propto R(t) \equiv  \sum_{n=1}^{\infty} \psi_{n}(t),
\end{equation}
where $\psi_{n}(t)$ is the probability that up to time $t$, $n$ recurrences occur, the last taking place exactly at time $t$. The assumption  that the times $\tau_{i}$ are uncorrelated yields $\hat \psi_{n}(u) = (\hat \psi_{1}(u))^{n}$. Note that $\psi_{1}(t)$ is the waiting time distribution function $\psi_{D}(t)$ of which we are trying to assess the asymptotic properties. Note also that for $u \rightarrow 0$ the Laplace transform of $\psi_{D}(t)$ with the form $\psi_{D}(t) = (\mu_{D} -1)\, T^{\mu_{D}-1}/(t+T)^{\mu_{D}}$ is \cite{klafter} $\hat \psi_{D}(u) = 1 - c\ u^{\mu_{D}-1}$, with $c = \Gamma(2-\mu_{D}) T^{\mu_{D} -1}$. The parameter $T<1$ is introduced to ensure the normalization condition. In the asymptotic limit of small $u$'s, $\hat R(u)$, which is $(1-c\, u^{\mu_{D} -1})/c\, u^{\mu_{D}-1} $, becomes $1/c\, u^{\mu_{D}-1}$. As shown in Ref.~\cite{mauro}, this is the limit for $u \rightarrow 0$ of the Laplace transform of a function of time, which for $t \rightarrow \infty$ is proportional to $1/t^{2-\mu_{D}}$. By comparison with Eq.~(\ref{particleatorigin}) we get the important relation of Eq.~(\ref{individualandsociety}).

\begin{figure}[ht]
\begin{center}
\includegraphics[height=2.1in,width=3.375in]{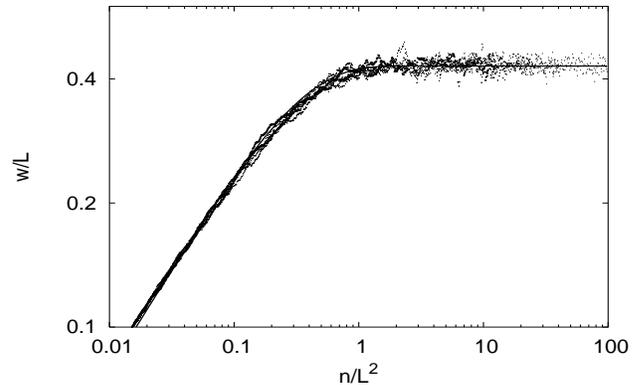}
 \caption{\label{fig:bianconi}Theoretical standard deviation of the diffusion process described by Eq.~(\ref{langevin}) (solid line), compared to collapsed numerical results
for the interface width from the "self-similar" model of Ref.~\cite{bianconi} (dots). Q and $\gamma$ were chosen to fit the data points.}
\end{center}
\end{figure}

The earlier remarks refer to the free diffusion part of the process. To derive the full picture afforded by the KPZ theory, free diffusion and saturation alike, we subordinate the growth process to  the ordinary Langevin equation
\begin{equation}
\label{langevin}
\frac{dy}{dn} = - \gamma\,y(n) + \eta(n),
\end{equation}
with the dissipation $\gamma$ fitting the condition $\gamma \ll 1$ and $\eta(n)$ being an ordinary Gaussian white noise, defined by
\begin{equation}
\left< \eta(n)\eta(n')\right> = 2\,Q\,\delta\left(|n-n'|\right),
\end{equation}
with  $Q$ denoting the noise intensity. The dimensionless time $1/\gamma \gg 1$ is essentially the saturation time of this ordinary fluctuation-dissipation process. Note that a theoretical model for the random growth of surfaces, called "self-similar"  model by the authors of Ref.~\cite{bianconi}, and proven by them to yield $\beta = 1/2$, is accurately described by the Langevin equation of Eq.~(\ref{langevin}), as seen in Fig.~\ref{fig:bianconi}. This means that  the interaction of the column under study with its two nearest neighbors is properly taken into account by the dissipation term, i.e.,  the first term on the right hand side of Eq.~(\ref{langevin}).

\begin{figure}[ht]
\begin{center}
\includegraphics[height=2.1in,width=3.375in]{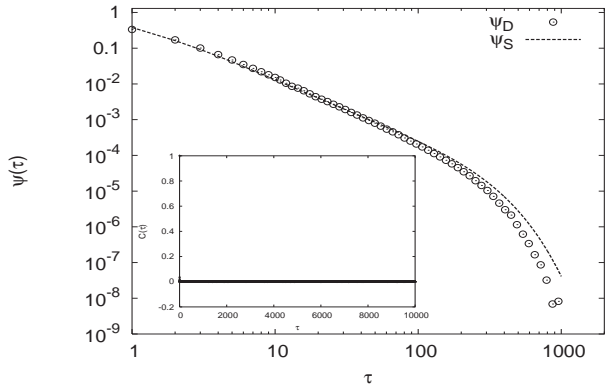}
\caption{The waiting time distribution $\psi(\tau)$ as a function of time $\tau$. The circles denote 
the waiting time distribution $\psi_{D}(\tau)$ of the BD model. The dashed line illustrates the theoretical waiting time distribution $\psi_{S}(\tau)$ used in Fig.~\ref{fig:relaxation} to reproduce the KPZ theory, through a proper use of the BD model, as illustrated in the text. The insert shows the correlation $C(\tau)$ of the waiting times $\tau_{i}$ as a function of time $\tau$ (from the BD model)} \label{fig:psioftauAndcorr}
\end{center}
\end{figure}

At this stage we have to determine numerically the form of $\psi_{D}(t)$ in the case of the BD model, which is known to fall in the basin of attraction of the KPZ equation. The results are illustrated in Fig.~\ref{fig:psioftauAndcorr}. We see that $\psi_{D}(t)$ is an inverse power law, with the expected index $\mu_{D} = 5/3$, truncated at times of the order of saturation times by a faster  decay.  It is important to notice that the result of the numerical calculation illustrated in the insert of Fig.~\ref{fig:psioftauAndcorr}, shows that the recurrence times $\tau_{i}$ are not correlated. These numerical results justify the identification of $\psi_{S}(\tau)$ with $\psi_{D}(\tau)$, and show that the subordination process must be realized with an inverse power law truncated at times of the order of saturation times by a faster decay. 

%
%
 The process of subordination, yielding both anomalous free diffusion and saturation, is realized through the following equation \cite{sokolov,metzlerandklafter,barkaiandsilbey}
\begin{equation}
\label{ordinary}
\frac{\partial p(y,t)}{\partial t} = \int_0^t \Phi(t-t')\,\gamma\,\left[\frac{d}{d y} y + \left< y^{2}\right>_{eq}\frac{d^{2}}{d y^{2}}\right]\,p(y,t')\, dt',
\end{equation}
 where $\Phi(t)$ is defined by means of its Laplace transform as
\begin{equation}
\label{definition}
\hat \Phi(u) = u\, \hat\psi_{S}(u)/(1 - \hat\psi_{S}(u))
\end{equation}
and $\left<y^{2}\right>_{eq} \equiv Q/\gamma$. Note that the subordination process has the effect of turning, in a statistical sense,  time $n$ into the real time $t = n^{1/\alpha}$. Thus, we have to turn the saturation time $1/\gamma$ of the ordinary fluctuation-dissipation process into the much larger value $1/\gamma^{1/\alpha} \gg 1/\gamma \gg 1$. At times much shorter than the saturation time $1/\gamma^{1/\alpha}$, $\psi_{S}(t)$ is an inverse power law with index $\mu_{S} = \mu_{D} =5/3$. This means that the Laplace transform of the memory kernel $\Phi(t)$ for $u \rightarrow 0$, for small values of $u$ fitting though the condition $u \gg \gamma^{1/\alpha}$, regime (i),   gets the same form as the free memory kernel $\Phi_{0}(t)$ used by the authors of Refs.\cite{sokolov,metzlerandklafter,barkaiandsilbey}, 
\begin{equation}\label{memorykernel}
\hat \Phi(u) \approx \hat \Phi_{0}(u) \equiv  \frac{1}{c}\,u^{(1-\alpha)}.
\end{equation}

For simplicity we set $c = 1$, namely, $T\equiv \left({1}/{\Gamma(1-\alpha)}\right)^{1/\alpha}$. 
Furthermore, we derive  $\psi_{S}(\tau)$ from Eq. (\ref{definition}) with the modified memory kernel
\begin{equation}\label{newmethod}
\Phi(t) = \Phi_{0} (t)\, \text{exp}(-\gamma^{1/\alpha} t),
\end{equation}
thereby setting $\hat \Phi(u)$ of Eq.(\ref{definition}) equal to $\hat \Phi_{0}(u+\gamma^{1/\alpha})$. This yields  $\hat \psi_{S}(u)=\hat \Phi_{0}(u+\gamma^{1/\alpha})/(u+\hat \Phi_{0}(u+\gamma^{1/\alpha}))$, whose Laplace transform determines $\psi_{S}(\tau)$.
The comparison with the numerical $\psi_{D}(\tau)$ requires some caution. In fact, the BD short-time behavior is model dependent and the subordination method is asymptotic in time. This requires that the comparison is done with a proper shift of $\psi_{S}(\tau)$. Nevertheless, after the shift, the decay of $\psi_{S}(\tau)$ turns out to be slightly slower than that of $\psi_{D}(\tau)$, an expected finite size effect \cite{stanley}.

We now prove that the memory kernel $\Phi(t)$, making $\psi_{S}(\tau) = \psi_{D}(\tau)$, compatibly with the limitation posed by the BD model as a substitute of the KPZ theory, satisfactorily reproduces the main
KPZ properties.  By using the Laplace transform approach we derive from Eq.~(\ref{ordinary}), with the  form for $\Phi(t)$ given by Eq. (\ref{newmethod}), the following analytical expressions for $\left< y(t) \right>$ and $\left< y^{2}(t) \right>$:
\begin{equation}
\label{onsager}
\left< y(t) \right> = \left< y(0) \right> K_{\alpha}^{(1)}(t)
\end{equation}
and
\begin{equation}
\label{impressive}
\left< y^{2}(t) \right> = \left< y^{2}(0) \right> K_{\alpha}^{(2)}(t) + \left< y^{2} \right>_{eq}\left(1-K_{\alpha}^{(2)}(t)\right).
\end{equation}
The relaxation functions $K_{\alpha}^{(1)}(t)$ and $K_{\alpha}^{(2)}(t)$ are defined through their Laplace transforms
\begin{equation}
\label{laplace1}
\hat K_{\alpha}^{(1)}(u)  = \frac{1}{u + \gamma(u+\gamma^{1/\alpha})^{1-\alpha}}
\end{equation}
and
\begin{equation}
\label{laplace2}
\hat K_{\alpha}^{(2)}(u)  = \frac{1}{u + 2\, \gamma\,(u+\gamma^{1/\alpha})^{1-\alpha}}.
\end{equation}

In regime (i) the functions $K^{(i)}_{\alpha}(t)$ are indistinguishable from the Mittag-Leffler functions that would result from setting $\Phi(t) = \Phi_{0}(t)$ \cite{barkaiandsilbey}. Furthermore, for even shorter times corresponding to $u >\gamma$, these Mittag-Leffler functions turn out \cite{barkaiandsilbey} to be identical to stretched exponentials. Consequently,  $K^{(2)}_{\alpha}(t) \simeq \text{exp}(- 2 \gamma t^{\alpha})$. Making $u$ larger and $t$ shorter, the standard deviation $w$ can be derived from Eq.~(\ref{impressive}) by the short-time Taylor series expansion of the function $K^{(2)}_{\alpha}(t)$. Thus we have  $w(t) \propto t^{\beta}$,
with $\beta$ given by Eq. (\ref{traditional}), in full accordance with the literature on the random growth of surfaces. 
Note that the choice of Eq.(\ref{newmethod}) annihilates the slow tail of the Mittag-Leffler function, thereby  locating the onset of saturation in the time region of $(1/\gamma)^{1/\alpha}$. This can be easily shown by considering regime (ii), $u \ll \gamma^{1/\alpha}$, which makes Eq.~(\ref{ordinary}) equivalent to the Fokker-Planck equation associated to Eq.~(\ref{langevin}), with $\gamma$ replaced by $\gamma^{1/\alpha}$.

\begin{figure}[ht]
\begin{center}
\includegraphics[height=2.1in,width=3.375in]{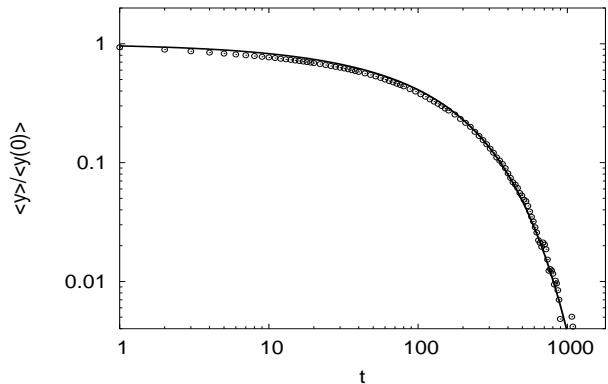}
\caption{\label{fig:relaxation}Regression of the out-of-equilibrium fluctuation $\left< y \right>$. The result of a $1 + 1$  dimensional BD simulation for a system of size L=200 (points) is compared to the regression form described by Eq.~(\ref{onsager}) (solid line), with $\gamma $ as the only fitting parameter,  and $\alpha=2/3$.}
\end{center} 
\end{figure}

We have to point out that in real systems a regime of transition to the steady state exists, this being of virtually vanishing time duration in the ordinary case of Fig.~\ref{fig:bianconi}. In the anomalous case this regime of transition becomes much more extended in time, thereby making it difficult to check in this case the theoretical predictions of Eqs.~(\ref{onsager}) and (\ref{impressive}). It is possible, however, to check the prediction of Eq.~(\ref{onsager}) by making an experiment at equilibrium, so as to avoid the out-of equilibrium induced  aging effects.  We let the system evolve till it reaches the steady state. Then, we  label all the columns whose height is larger than the average height. We observe the time evolution of only the labelled columns and we make an average on many identically prepared samples. The result is shown in Fig.~\ref{fig:relaxation}, and the accordance between theory and experiment is remarkable.

Note that according to the current literature, the standard deviation $w(L,t)$, with $L$ being the sample size,  
obeys the following form of scaling
\begin{equation}
w(L,t) =  L^{1/2} f\left( {\frac{t^{2\beta}}{L}} \right).
\end{equation}
This is valid for both the KPZ and the Edward-Wilkinson \cite{stanley} theory.  $f(x)$ is an unknown function fitting the conditions $f(x)\propto x^{\beta}$ if $x\ll 1$ and $f(x) = \text{const} $ if $x\gg 1$. The theory of this paper determines $f(x)$ as follows. We identify $L$ with $1/\gamma$, thereby setting $x = t^{2\beta} \gamma$. Then, we note that the KPZ condition  $\beta = 1/3$ is obtained from the constraint that $\mu_{D}$ of Eq.~(\ref{individualandsociety}) is identical to  $\mu_{S}$ of Eq.~(\ref{traditional}). In this condition the theory of this paper assigns to the unknown function $f(x)$ the following expression
\begin{equation}
\label{mainresult}
f(x) = \left(1 - K^{(2)}_{\frac{2}{3}}(x)\right)^{\frac{1}{2}}.
\end{equation}
To the best of our knowledge,  the subordination perspective is adopted here for the first time to account for the behavior of a single individual of the KPZ system. This is realized, on the other hand, with a proper choice of $\psi_{S}(\tau)$,  which  is compatible with the transition to the Markov condition in the long-time limit. 

Due to the random choice of the columns \cite{stanley}, the time distance between two consecutive arrivals  of particles in the same column must be Poisson. How to convert this Poisson-like phenomenon into an
anomalous subdiffusion, so as to involve the subordination method? The numerical results suggest that most of the arrival events are pseudo-events, which do not afford significant contributions to the spreading of the diffusion distribution  \cite{memorybeyondmemory}. Only some of these arrival events do, and the subordination function $\psi_{S}(\tau)$ is the distribution of time distances between two consecutive ones of them.

Finally, we want to express some conjectures about the role of correlated noise \cite{correlatednoise}.  
These authors \cite{correlatednoise}  proved that spatial and time correlation can make the scaling exponent $\beta$ of the EW model get values larger than $\beta = 1/4$.  What about the subordination approach of this paper and the models with correlated noise, either EW or KPZ, in the $d+1$ as well as $1+1$-dimensional case?  It is expected that the time distances between two consecutive random events, and with it the index $\mu_{S}$, are affected by both dimension and correlation. The anomalous coefficient $\beta$ is determined by $\beta = (\mu_{S}-1)/2$ (Eq.{\ref{traditional}). Thus, in principle, it is possible to recover the value of $\beta$ predicted by Ref. \cite{correlatednoise}, provided this is smaller than $1/2$. Thus, the $1+1$ EW, with its $\beta = 1/4$ can certainly be reproduced by means of our subordination approach. Of course, the condition $\mu_{S} = \mu_{D}$ would be lost. This seems to be an interesting conclusion, since it suggests further research work to do to assess why $\mu_{S} = \mu_{D}$  corresponds to the $1+1$ BD, which falls in the basin of attraction of the $1+1$ KPZ theory with uncorrelated noise.
 
We acknowledge financial support from ARO through Grant DAAD19-02-1-0037.


\begin{thebibliography}{100}

\bibitem{KPZ} M. Kardar, G. Parisi and Y.-C. Zhang, Phys. Rev. Lett. {\bf 56}, 889 (1986).
\bibitem{zhang} T. Halpin-Healy and Y.-C. Zhang, Phys. Rep. {\bf 254}, 215 (1995). 
\bibitem{stanley} A.-L. Barab\'asi, \emph{Fractal Concepts in Surface Growth}, (Cambridge University Press, Cambridge, 1995). 
\bibitem{meakin} P. Meakin, \emph{Fractals, scaling and growth far from equilibrium}, (Cambridge University Press, Cambridge, 1998).
\bibitem{moro} D. B. Abraham, R. Cuerno, and E. Moro, Phys. Rev. Lett. {\bf 88}, 206101 (2002). 
\bibitem{more} T. Karabacak, J. P. Singh, Y.-P. Zhao, G.-C. Wang, and T.-M. Lu, Phys. Rev. B {\bf 68}, 125408 (2003).
\bibitem{family} F. Family, Physica A {\bf 168} 561 (1990).
\bibitem{bak} P. Bak, C. Tang and K. Wiesenfeld, Phys. Rev. Lett. {\bf 59}, 381 (1987). 
\bibitem{chow} T.S. Chow, Phys. Rev. Lett. {\bf 79}, 1086 (1997). 
\bibitem{merikoski} J. Merikoski, J. Maunuksela, M. Myllys, J. Timonen and M. J. Alava, Phys. Rev. Lett. {\bf90}, 024501 (2003). 
\bibitem{dingdang} M. Ding and W. Yang, Phys. Rev. E {\bf 52}, 207 (1995). 
\bibitem{krugandobbs} J. Krug and H.T. Dobbs, Phys. Rev. Lett. {\bf 76}, 4096 (1996). 
\bibitem{searson} P.C. Searson, R. Li, and K. Sieradzki, Phys. Rev. Lett. {\bf 74}, 1395 (1995).
\bibitem{questioningthemsleves} J. Krug, H. Kallabis, S.N. Majumdar, S.J. Cornell, A. J. Bray, and C. Sire, Phys. Rev. E {\bf 56}, 2702 (1997). 
\bibitem{majumdar} S. N. Majumdar, Phys. Rev. E {\bf 68}, 050101 (R) (2003). 
\bibitem{sokolov} I.M. Sokolov, Phys. Rev. E {\bf 66}, 041101 (2002).
\bibitem{barkaiandsilbey} E. Barkai and R. J. Silbey, J. Phys. Chem. B, {\bf 104}, 3866 (2000). 
\bibitem{metzlerandklafter} R. Metzler and J. Klafter, J. Phys. Chem. B, {\bf 104}, 3851 (2000).
\bibitem{montrollandweiss} E.W. Montroll and G.H. Weiss, J. Math. Phys. {\bf 6}, 178 (1965). 
\bibitem{klafter} G. Zumofen and J. Klafter, Phys. Rev. E {\bf 47}, 851 (1993). 
\bibitem{mauro} M. Bologna, P. Grigolini, and B.J. West, Chem. Phys. {\bf 284}, 115 (2002). 
\bibitem{bianconi} G. Bianconi, M. A. Munoz, A. Gabrielli, and L. Pietronero, Phys. Rev. E {\bf 60}, 3719 (1999). 
\bibitem{memorybeyondmemory} P. Allegrini, P. Grigolini, P. Hamilton, L. Palatella, and G. Raffaelli, Phys. Rev. E {\bf 65}, 041926 (2002). 
\bibitem{correlatednoise} Yi-Kuo Yu and Ning-Ning Pang, T. Halpin-Healy, Phys. Rev. E {\bf 50}, 5111 (1994). 


\end{thebibliography}
\end{document}